\documentclass[preprint,psfig]{aastex}

\def\ltsima{$\; \buildrel < \over \sim \;$}

\def\lsim{\lower.5ex\hbox{\ltsima}}
\def\gtsima{$\; \buildrel > \over \sim \;$}
\def\gsim{\lower.5ex\hbox{\gtsima}}

\begin{document}
\title{A Physical Model for the Luminosity Function of High-Redshift
Quasars}

\author{J. Stuart B. Wyithe\altaffilmark{1} and Abraham Loeb}

\affil{Harvard-Smithsonian Center for Astrophysics, 60 Garden St.,
Cambridge, MA 02138}
\email{swyithe@cfa.harvard.edu; aloeb@cfa.harvard.edu}

\altaffiltext{1}{Hubble Fellow}

\begin{abstract}
\noindent 

We provide a simple theoretical model for the quasar luminosity function at
high redshifts that naturally reproduces the statistical properties of the
luminous SDSS quasar sample at redshifts $z\sim4.3$ and $z\ga 5.7$. Our
model is based on the assumptions that quasar emission is triggered by
galaxy mergers, and that the black hole mass is proportional to a power-law
in the circular velocity of the host galactic halo, $v_c$. We assume that quasars 
shine at their Eddington
luminosity over a time proportional to the mass ratio between the small and
final galaxies in the merger.  This simple model fits the quasar luminosity
function at $z\sim2$--$3$, reproduces the normalization and logarithmic
slope ($\beta\sim-2.58$) at $z\sim4.3$, explains the space density of
bright SDSS quasars at $z\sim6.0$, reproduces the black hole -- halo mass
relation for dormant black holes in the local universe, and matches the
estimated duty cycle of quasar activity ($\sim 10^7$ years) in Lyman-break
galaxies at $z\sim 3$.  Based on the derived luminosity function we predict
the resulting gravitational lensing rates for high redshift quasars.  The
lens fractions in the SDSS samples are predicted to be $\sim 2\%$ at
$z\sim4.3$ and $\sim 10\%$ at $z\ga5.7$.  Interestingly, the limiting
quasar luminosity in our best-fit relation $L \propto v_c^5/G$, scales
as the binding energy of the host galaxy divided by its dynamical time,
implying that feedback is the mechanism that regulates black hole growth in
galactic potential wells.

\end{abstract}


\section{Introduction}

While the quasar luminosity function has been studied extensively at
redshifts below $z\sim3$ (e.g. Boyle, Shanks \& Peterson~1988; Hartwick \&
Schade 1990; Pei~1995), the Sloan Digital Sky Survey (SDSS; Fukugita et
al.~1996; Gunn et al.~1998; York et al.~2001) has in recent years
substantially increased the number of quasars known at $z\ga3.5$ (Fan,
Strauss et al.~2001a,b; Schneider et al.~2001). Two samples of very high
redshift SDSS quasars have been presented to date. The first of these is a
sample of 39 luminous quasars with redshifts in the range $3.6<z<5.0$ and a
median of 4.3. The second sample consists of 4 quasars at $z\ga5.7$,
including the quasar with the highest known redshift ($z=6.28$).  These
samples begin to sketch out the luminosity function of quasars at $z\ga4$
and are very important for studies of the ionizing background radiation and
of quasar evolution around the epoch of reionization (Haiman \&
Loeb~1998). Although the logarithmic slope of the luminosity function for
bright quasars admits a universal value of $\sim-3.5$ at redshifts $z\la
3$, Fan et al.~(2000a) find a shallower logarithmic slope of $\sim-2.5$ for
their luminous quasars at $z\ga 3.5$. At first sight this is a puzzling
result because the exponential tail of the Press-Schechter~(1974) mass
function describing the space density of the massive host galaxies in which
bright quasars are thought to reside, becomes progressively steeper with
increasing redshift.
 
There are several published models of varying complexity to describe the
observed evolution of the quasar luminosity function within hierarchical
structure formation theory (Efstathiou \& Rees 1988; Small \& Blandford
1992; Haehnelt \& Rees 1993; Haiman \& Loeb~1998; Haehnelt Natarajan \&
Rees~1998; Kauffmann \& Haehnelt~2000).  In these models, the luminosities
and lifetimes of the quasars are governed by the black hole mass and the
supply of cold accreting gas. The abundance and evolution of the
supermassive black holes assumed to power the quasars, are linked to the
evolution of the mass-function of galactic halos.  The density of bright
quasars is observed to decline rapidly with redshift below $z\sim2$ (Boyle,
Shanks \& Peterson~1988; Hartwick \& Schade 1990). This decline has been
explained (Kauffman \& Haehnelt~2000; see also Cavaliere \& Vittorini~2000)
in terms of decreases in the merger rate and the availability of cold gas
to fuel the black holes.  In addition, the density of bright quasars is
also observed to decline beyond a redshift of $z\sim2$ (Warren, Hewett \&
Osmer~1994; Schmidt, Schneider \& Gunn~1995; Kennefick, Djorgovski \& de
Carvalho~1995). Under the assumption that quasar activity is related to the
formation rate of halos, model luminosity functions have been constructed
that successfully describe this decline and the shape of the luminosity
function between $z\sim2$ and $z\sim3.5$ (Efstathiou \& Rees~1988; Haehnelt
\& Rees~1993; Haiman \& Loeb~1998; Haehnelt, Natarajan \& Rees~1999;
Haiman, Madau \& Loeb~1999). These models predict the luminosity function
at still higher redshifts. With the recent influx of results on the quasar
luminosity at redshifts above $z\sim4$, it is interesting to revisit the
question of modeling the evolution of the quasar luminosity function at
high redshifts. In this paper, we demonstrate that a very simple model can
explain the new results on the high redshift quasar luminosity function.

Our model for the evolution of the quasar luminosity function is based on
the Press-Schechter~(1974) mass function and halo merger rates computed
with the excursion set formalism (Bond et al.~1991; Lacy \& Cole~1993). We
show that this model can reproduce all of the known properties of the
luminosity function above $z\sim2$. We begin in \S\ref{models} by
describing an approximate version of our model (Haiman \& Loeb~1998;
hereafter HL98), followed by the full discussion based on the halo merger
rates. In \S\ref{lumfuncobs} and \S\ref{cp} we discuss the data on the
quasar luminosity function and its comparison with our model.  Finally we
discuss the implications of our model for gravitational lensing in
high-redshift samples of quasars in \S\ref{lens}, and present our main 
conclusions in \S\ref{conclusion}. Throughout the paper we assume density
parameters of $\Omega_{m}=0.35$ in matter, $\Omega_\Lambda=0.65$ in a
cosmological constant, and a Hubble constant of $H_0=65~{\rm
km\,s^{-1}\,Mpc^{-1}}$. For calculations of the Press-Schechter~(1974)
mass function we assume a primordial power-spectrum with $n=1$ and the
fitting formula to the exact transfer function of Bardeen et al.~(1986).

\section{Theoretical Luminosity Functions}
\label{models}

We compare two models for the quasar luminosity function that are based on
the Press-Schechter~(1974) mass function of dark matter halos, and the halo
merger rates from the excursion set formalism of Bond et al.~(1991) and
Lacy \& Cole~(1993).  We begin with the approximate model of HL98 which is
based on the halo formation rate.  We then describe our more complete model
based on halo merger rates. We will demonstrate in \S\ref{comp} that only this
latter model reproduces all the known features of the quasar luminosity
function above $z\sim2$, as well as the local black hole -- host halo mass
relation and new constraints on the quasar duty cycle.

\subsection{An Approximate Model for the Quasar Luminosity Function 
Based on the Halo Formation Rate}
\label{modelhl}

HL98 discussed an approximate model for the high redshift quasar luminosity
function by supposing that quasars are associated with newly formed
halos. Their model assumes that each galactic halo hosts a black hole with
a mass ($M_{\rm bh}$) which is a constant fraction ($\epsilon$) of the host
halo mass ($M_{\rm halo}$), and that the black hole shines at its Eddington
rate with a universal light-curve, $f(t)$.
HL98 used the derivative of the Press-Schechter mass function to
approximate the formation rate of halos, arguing that mergers are rare at
high redshift so that the negative contribution to the derivative is
negligible.  In this model the quasar luminosity as a function of time is
\begin{equation}
L_{\rm B}(t)=M_{\rm bh}f(t)=\epsilon M_{\rm
halo}f(t)\hspace{5mm}\mbox{for}\hspace{5mm}M_{\rm halo}>M_{\rm min},
\end{equation}
where $M_{\rm min}\sim10^8M_{\rm \odot}[(1+z)/10]^{-\frac{3}{2}}$ is the
minimum halo mass inside which a black hole can form.  This lower mass
limit corresponds to the virial temperature below which atomic cooling is
not effective in allowing the gas to sink to the center (Barkana \& Loeb
2001). Following HL98 we write the resulting quasar luminosity function,
defined as the comoving number density of quasars having rest frame B-band 
luminosities between $L_{\rm B}$ and $L_{\rm B}+\Delta L_{\rm B}$ and redshifts 
between $z$ and $z+\Delta z$, as
\begin{equation}
\label{LF1}
\Psi_{\rm HL}(L_{\rm B},z) = \int_z^\infty dz'\int_{\rm \epsilon M_{\rm
min}}^\infty dM_{\rm bh}\frac{d^2n_{\rm bh}}{dM_{\rm bh}dz'}\delta[L_{\rm B}-M_{\rm
bh}f(t_{\rm z}-t')],
\end{equation}
where $t_{\rm z}$ and $t'$ are the time at redshifts $z$ and $z'$, and
$\frac{d^2n_{\rm bh}}{dM_{\rm bh}dz'}$ is the change in the co-moving
number density of black holes between $z'$ and $z'+dz'$. Integrating over
$M_{\rm bh}$ we get
\begin{equation} 
\Psi_{\rm HL}(L_{\rm B},z) = \int_{\rm
0}^{a=\frac{1}{1+z}}da {dz \over da}\frac{1}{f(t_{\rm
z}-t)}\left.\frac{d^2n_{\rm bh}}{dM_{\rm bh}dz}\right|_{\rm M_{\rm
bh}=\frac{L_{\rm B}}{f(t_{\rm z}-t)}} ,
\end{equation}
where $a=(1+z)^{-1}$ is the scale factor at a redshift $z$. HL98 used an
exponential with an $e$-folding time $t_{\rm dc,0}$ for the universal light
curve $f$. We find that the equivalent result for $\Psi_{\rm HL}(L_{\rm B},z)$ is
obtained more simply by assuming a step function for $f$, namely
\begin{equation}
\label{lcurve}
f(t)=\frac{L_{\rm Edd,B}}{M_{\rm bh}}\Theta(t-t_{\rm dc,0}), 
\end{equation} 
where $L_{\rm Edd,B}=5.7\times10^{3}({M_{\rm bh}}/{M_\odot})$ is the
Eddington luminosity in B-band solar luminosity units for the median quasar
spectrum (Elvis et al.~1994), and $\Theta(t)$ is the Heaviside step
function.  We then find
\begin{equation} 
\label{phiHL5}
\Psi_{\rm HL}(L_{\rm B},z) = \int_{\rm a_{\rm min}=\frac{1}{1+z_{\rm
form}}}^{a=\frac{1}{1+z}}da\frac{dz}{da}\frac{M_{\rm bh}}{L_{\rm
Edd,B}}\left.\frac{d^2n_{\rm bh}}{dM_{\rm bh}dz}\right|_{\rm M_{\rm
bh}=\frac{L_B}{f(t_{\rm z}-t)}}
\end{equation}
given the formation redshift for the halo, $z_{\rm form}$. Assuming that
$t_{\rm dc,0}\ll H^{-1}(z)$, $(a-a_{\rm min})\approx ({da}/{dt})t_{\rm
dc,0}$, and relating the space density of black holes to the
Press-Schechter~(1974) mass function through
\begin{equation}
\frac{d^2n_{\rm bh}}{dM_{\rm bh}dz}=\frac{1}{\epsilon}\frac{d^2n_{\rm ps}}{dM_{\rm halo}dz},
\end{equation}
we obtain the following simple expression for the quasar luminosity
function:
\begin{equation} 
\Psi_{\rm HL}(L_{\rm B},z) \approx \frac{t_{\rm dc, 0}}{H_0^{-1}}
\sqrt{{\Omega_m\over a^5}+{\Omega_{\rm
\Lambda}\over a^2}}\frac{1}{5.7\times10^{3}}
\frac{1}{\epsilon}\left.\frac{d^2n_{\rm ps}}{dM_{\rm halo}dz}\right|_{\rm
M_{\rm halo}=\frac{L_B}{5.7\times10^3\epsilon}}
\end{equation}
We will compare this model to observations in \S\ref{compHL}.

\subsection{A Model for the Quasar Luminosity Function Based on the 
Halo Merger Rate}
\label{lumfunc}

The HL98 model associates quasar activity with halo formation.  The
assumption is that the halo merger rate is rarer than the halo formation
rate at high redshift. However, at low redshifts the formation rate becomes
smaller than the merger rate, and the derivative of the Press-Schechter
mass function on which the model is based, becomes negative. Quasar
activity at low redshift can be explained in relation to halo merger
activity (Carlberg~1990; Kauffmann \& Haehnelt~2000). In this section we
discuss a simple model that associates quasar activity at high redshift
with the merger rate of halos.

The merger rate of halos was computed by Lacy \& Cole~(1993) based on the
excursion set formalism of Bond et al.~(1991). We compute the number of
halos having masses between $\Delta M_{\rm halo}$ and $\Delta M_{\rm halo}
+ d\Delta M_{\rm halo}$ that accrete onto a halo of mass $M_{\rm
halo}-\Delta M_{\rm halo}$ per unit time $\left.\frac{d^2N_{\rm
merge}}{d\Delta M_{\rm halo} dt}\right|_{\rm M_{\rm halo}-\Delta M_{\rm
halo}}$.  
Thus, the number of merger events involving the accretion of a halo of mass
$\Delta M_{\rm halo}$ by a halo of mass $M_{\rm halo}-\Delta M_{\rm halo}$
per unit time per comoving volume at a redshift $z$ is given by the product
\begin{equation}
\left.\frac{dn_{\rm ps}}{dM}\right|_{\rm M=M_{\rm halo}-\Delta M_{\rm
halo}}\times\left.\frac{d^2N_{\rm merge}}{d\Delta M_{\rm halo}
dt}\right|_{\rm M_{\rm halo}-\Delta M_{\rm halo}}.
\end{equation}

We assume that the mass of the central black hole scales as a power-law
with the circular velocity ($v_{\rm c}$) of the halo, 
\begin{equation}
M_{\rm bh} \propto v_{\rm c}^\gamma .
\end{equation} 
The circular velocity of a halo of mass $M_{\rm halo}$ at redshift $z$ 
can be written as (Barkana \& Loeb~2001),
\begin{equation}
\label{vc}
v_{\rm c} = 159.4 \left(\frac{M_{\rm halo}}{10^{12}
h^{-1}M_{\odot}}\right)^\frac{1}{3}
\left[\frac{\Omega_m}{\Omega_{m}^{z}}
\frac{\Delta_c}{18\pi^2}\right]^\frac{1}{6}
\left({1+z}\right)^\frac{1}{2}\mbox{km}\,\mbox{s}^{-1},
\end{equation}
where $h=(H_0/100~{\rm km~s^{-1}~Mpc^{-1}})$, $\Delta_{\rm c} =
18\pi^2+82d-39d^2$ is the final overdensity relative to the critical
density at redshift $z$, and $d\equiv1-\Omega_{m}^{z}$ with
$\Omega_{m}^{z}=\frac{\Omega_{m}(1+z)^3}{\Omega_{m}(1+z)^3
+ \Omega_\Lambda}$. We can therefore write,
\begin{equation}
\label{eps}
\frac{M_{\rm bh}}{M_{\rm halo}} = \epsilon = \epsilon_{\rm o}
\left(\frac{M_{\rm halo}}{10^{12}
M_{\odot}}\right)^{\frac{\gamma}{3}-1}\left[\frac{\Omega_m}{\Omega_{m}^{z}}\frac{\Delta_c}{18\pi^2}\right]^\frac{\gamma}{6}h^{\gamma/3}(1+z)^\frac{\gamma}{2},
\end{equation}
and fit for two free parameters $\epsilon_{\rm o}$ and $\gamma$ that
describe the evolution and slope of the black hole -- halo mass
relation. We assume that the black holes coalesce upon halo merger
(Kauffmann \& Haehnelt~2000) and find the number of black holes of mass
between $\Delta M_{\rm bh}$ and $\Delta M_{\rm bh} + d\Delta M_{\rm bh}$
that merge with black holes of mass $M_{\rm bh}-\Delta M_{\rm bh}$ per unit
time:
\begin{equation}
\left.\frac{d^2N_{\rm merge}}{d\Delta M_{\rm bh} dt}\right|_{\rm M_{\rm
bh}-\Delta M_{\rm bh}}=\frac{3}{\gamma\epsilon}\left.\frac{d^2N_{\rm
merge}}{d\Delta M_{\rm halo} dt}\right|_{\rm M_{\rm halo}-\Delta M_{\rm
halo}}.
\end{equation}  
A linear relation ($\gamma=3$) between black hole mass and halo mass is a
natural consequence of black hole growth that is dominated by coalescence
with no gas accretion (Haehnelt et al. 1998). Values of $\gamma>1$ result
from significant gas accretion during the active quasar phase as deduced by
Yu \& Tremaine~(2002). We assume that after a merger, a fraction of the
cold gas from the accreted halo is driven onto the central black hole of
mass $M_{\rm bh}$ (Mihos \& Hernquist~1994; Hernquist \& Mihos~1995). If
the quasar shines at the Eddington rate of the black hole in the merger
product (see, e.g. Yu \& Tremaine~2002), then the cold gas from the small
accreted halo (which makes up the new fuel reservoir) will run out in a
time approximately proportional to
\begin{equation}
\label{timeratio}
\frac{\Delta M_{\rm baryon}}{M_{\rm halo}} = \frac{\Delta M_{\rm
halo}}{M_{\rm halo}}\frac{\Delta M_{\rm baryon}}{\Delta M_{\rm halo}},
\end{equation}
where $\Delta M_{\rm baryon}$ is the mass in baryons within the accreted
halo\footnote{Equation~(\ref{timeratio}) can be derived as a Taylor expansion
for minor mergers. For simplicity we apply it to all mergers.}.  
Thus we postulate that after a merger the 
quasar shines at the Eddington rate corresponding to the merger product, 
for a time proportional
to both the ratio between the masses of the accreted and initial halo and
the baryon fraction of the accreted halo. 

\begin{figure*}[htbp]
\epsscale{.5}
\plotone{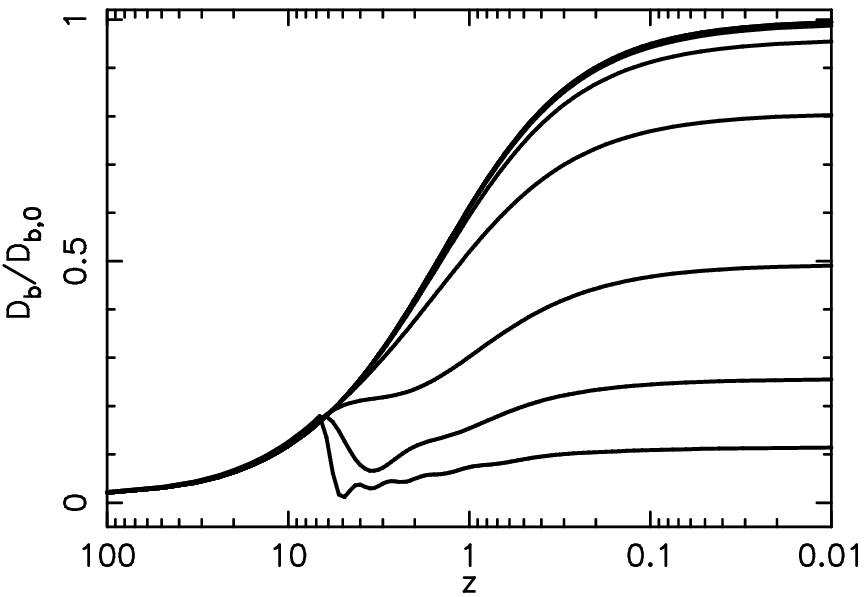}
\caption{\label{fig1} The growth factor $D_{\rm b}$ (normalized to the 
present day $D_{\rm b,0})$ for baryons associated with dark
matter halos of mass $M_{\rm halo}=10^6$, $10^7$, $10^8$,
$10^9$, $10^{10}$, $10^{11}$, $10^{12}$, $10^{13}$ and $10^{14}M_{\rm
\odot}$ (bottom to top).  The down turn of the growth factor for small
halos corresponds to the reionization epoch at $z_{\rm reion}\sim7.0$.}
\end{figure*}

In order to examine whether the baryon mass fraction in the accreted halo
would be affected by the photo-ionization heating of the intergalactic
medium (IGM) after reionization, we have solved the linear growth factors
$D_{\rm dm}$ and $D_{\rm b}$ for the dark matter and baryons on a spatial
comoving scale $R=\left({3\Delta M_{\rm halo}}/{4\pi\rho_{\rm
m}}\right)^{1/3}$, where $\rho_{\rm m}$ is the average matter density of
the universe today.
These growth factors obey the following set of coupled differential
equations for the dark matter and baryon overdensities $\delta_{\rm dm}$
and $\delta_{\rm b}$ in the linear regime (e.g. Barkana \& Loeb~2001):
\begin{eqnarray}	
\label{growthfac}
\nonumber &&\hspace{-15mm}\ddot{\delta}_{\rm dm} +2 H\dot{\delta}_{\rm dm}
= \frac{3}{2}H^2\left[\Omega_{\rm b}(z) \delta_{\rm b}+\Omega_{\rm
dm}(z)\delta_{\rm dm}\right]\\ &&\hspace{-15mm}\ddot{\delta}_{\rm b} +2
H\dot{\delta}_{\rm b} = \frac{3}{2}H^2\left\{[\Omega_{\rm b}(z) \delta_{\rm
b}+\Omega_{\rm dm}(z)\delta_{\rm dm}]\right\} - \frac{k_{\rm B} T_{\rm i}
}{\mu m_{\rm p}} \left(\frac{k}{a}\right)^2 \left(\frac{a_{\rm
i}}{a}\right)^{1+\beta} \left[\delta_{\rm b}+\frac{2}{3} \beta (\delta_{\rm
b}-\delta_{\rm b,i})\right] .
\end{eqnarray}
Here $T$ is the gas temperature, $\mu$ the mean molecular weight
($\sim0.69$ for ionized IGM), $k_{\rm B}$ is Boltzmans constant, and $k$ is
the co-moving wave number for the mode of interest.  The subscripts $i$
refer to quantities at an initial reference time and $\beta=0$ or 1 for
adiabatic or isothermal evolution. We assume reionization at $z_{\rm
reion}=7$. Before reionization we assume $T_{\rm i}=0$; the baryon and
dark-matter over-densities undergo the same evolution during this period.
Following reionization, the IGM is assumed to be heated to $10^4$K and to
evolve isothermally ($\beta=1$).  The evolution of $D_{\rm b}$ is described
by the superposition of solutions to equation~(\ref{growthfac}) weighted by
the Fourier transform of the spatial tophat window function of width $R$
(Medvigy \& Loeb~2001). Figure~\ref{fig1} shows the evolution of $D_{\rm
b}$ with redshift. The curves shown correspond (from bottom to top) to halo
masses of $M_{\rm halo}=10^6$, $10^7$, $10^8$, $10^9$, $10^{10}$,
$10^{11}$, $10^{12}$, $10^{13}$ and $10^{14}M_{\rm \odot}$. The dark matter
growth factor corresponds to the upper envelope in this figure. Prior to
reionization $D_{\rm b}$ follows $D_{\rm dm}$ for all halo
masses. Reinonizaton heats the IGM and eliminates gas accretion onto small
halos. The baryon overdensity in halos larger than $10^{10}M_{\rm \odot}$
is unaffected by reionization. Thus, we find that the effect of
reionization on the luminosity function is only apparent for quasars
fainter than $\sim10^{11}L_{\rm \odot}$, which is well below the detection
threshold of existing quasar surveys at $z\gg 1$, such as SDSS.  We
therefore assume a fixed baryon mass fraction in objects relevant for the
luminosity function of bright quasars at high redshifts.

The quasar light curve and luminosity function may therefore be written as
\begin{equation}
f(t)=\frac{L_{\rm Edd,B}}{M_{\rm bh}}\Theta\left(t-\frac{\Delta M_{\rm
halo}}{M_{\rm halo}} t_{\rm dc,0}\right),
\end{equation} 
and
\begin{eqnarray}
\label{LF1}
&&\hspace{-8mm}\Psi(L_{\rm B},z) = \\ \nonumber &&\hspace{-8mm}\int_{\rm \epsilon
M_{\rm min}}^\infty dM_{\rm bh} \int_{0}^{0.5 M_{\rm bh}} d\Delta
M_{\rm bh} \int_z^\infty dz' \left.\frac{dn_{\rm bh}}{dM}\right|_{\rm
M=M_{\rm bh}-\Delta M_{\rm bh}}\left.\frac{d^2N_{\rm merge}}{d\Delta M_{\rm
bh}dt'}\right|_{\rm M_{\rm bh}-\Delta M_{\rm bh}} \frac{dt'}{dz'}
\delta\left(L_{\rm B}-M_{\rm bh}f(t_{\rm z}-t')\right),
\end{eqnarray}
in analogy with equations~(\ref{lcurve}) and (\ref{phiHL5}),
respectively. Integrating over $M_{\rm bh}$ we find
\begin{eqnarray} 
&&\hspace{-8mm}\Psi(L_{\rm B},z) = \\ \nonumber &&\hspace{-8mm}\int_{\rm
0}^{0.5M_{\rm halo}} d\Delta M_{\rm halo} \int_{\rm a_{\rm
min}=\frac{1}{1+z_{\rm form}}}^{a=\frac{1}{1+z}}da\frac{dz}{da}
\left.\frac{dn_{\rm ps}}{dM}\right|_{\rm M=M_{\rm halo}-\Delta M_{\rm
halo}}\left.\frac{d^2N_{\rm merge}}{d\Delta M_{\rm halo}dt}\right|_{\rm
M_{\rm halo}-\Delta M_{\rm halo}}
\frac{3}{\gamma\epsilon}\frac{dt}{dz}\frac{M_{\rm bh}}{L_{\rm Edd,B}}.
\end{eqnarray}
Assuming $t_{\rm dc,0}\ll H^{-1}(z)$, we then obtain $a-a_{\rm
min}=\frac{da}{dt}t_{\rm dc,0}\frac{\Delta M_{\rm halo}}{M_{\rm
halo}}$, yielding the luminosity function
\begin{eqnarray}
\nonumber
\hspace{-8mm}&&\Psi(L_{\rm B},z)=\int_{0}^{0.5 M_{\rm halo}}d\Delta M_{\rm
halo}\frac{3}{\gamma\epsilon}\frac{t_{\rm dc,0}}{5.7\times10^3}\frac{\Delta M_{\rm
halo}}{M_{\rm halo}}\left.\frac{dn_{\rm
ps}}{dM}\right|_{\rm M=M_{\rm halo}-\Delta M_{\rm
halo}}\left.\frac{d^2N_{\rm merge}}{d\Delta M_{\rm halo}dt}\right|_{\rm
M_{\rm halo}-\Delta M_{\rm halo}}\\
\hspace{-8mm}&&\mbox{where}\hspace{2mm}M_{\rm halo}={L_B}/{
5.7\times10^3\epsilon}.
\end{eqnarray}
The value of $t_{\rm dc,0}$ can be related to the duty cycle of quasars
$t_{\rm dc}$ through consideration of the number of mergers during a Hubble
time $H^{-1}(z)$,
\begin{equation}
\label{dc}
t_{\rm dc}(L_{\rm B},z) = t_{\rm dc,0} \int_{0}^{M_{\rm halo}=\frac{0.5
L_B}{5.7\times10^3\epsilon}}d\Delta M_{\rm halo}\frac{\Delta M_{\rm
halo}}{M_{\rm halo}} H^{-1}(z) 
\left.\frac{d^2N_{\rm merge}}{d\Delta M_{\rm
halo}dt}\right|_{\rm M_{\rm halo}-\Delta M_{\rm halo}}
\end{equation}	
We will compare the luminosity function predicted by this model with
observations in \S\ref{comp}.

\section{The Observed Luminosity Function}
\label{lumfuncobs}

The standard double power-law luminosity function (Boyle, Shanks \&
Peterson~1988; Pei~1995)
\begin{equation}
\label{LF}
\Psi(L_{\rm B},z)=\frac{\Psi_*/L_*(z)}{[L_{\rm B}/L_*(z)]^{\beta_{\rm l}}+[L_{\rm B}/L_*(z)]^{\beta_{\rm h}}}
\end{equation}
provides a good representation of the observed quasar luminosity function
at redshifts $z\la 3$. At the faint end of the luminosity function, the
slope is $\beta_{\rm l}=1.51$, while at the bright end $\beta_{\rm
h}=3.43$. Moreover, all dependence on redshift (at $z\la3$) is in the break
luminosity $L_*$ indicating pure luminosity evolution. At low redshift the
break luminosity evolves as a power-law in redshift, and the space density
of bright quasars increases with redshift. However, surveys at higher
redshift show a decline in the space density of bright quasars beyond
$z\sim3$ (Warren, Hewett \& Osmer~1994; Schmidt, Schneider \& Gunn~1995;
Kennefick, Djorgovski \& de Carvalho~1995).

While quasar evolution below $z\sim3$ is described by pure luminosity
evolution, Fan et al.~(2001a) found from their sample of SDSS quasars at
$z\sim4.3$ that the slope at the bright end of the luminosity function has
evolved from the $z\la 3$ value of $\beta_{\rm h}\sim-3.5$ to $\beta_{\rm
h}\sim-2.5$. This result is supported by the analysis of Schmidt, Schneider
\& Gunn~(1995). Fan et al.~(2001a) also find an evolution of space density
with redshift for bright quasars of $\Psi\propto 10^{-0.5z}$ 
between $z\sim3.5$ and $z\sim5$. The space density of bright quasars
measured by Fan et al.~(2001b) at $z\sim6$ agrees with the extrapolation of
this evolution.

We seek to model the observed features of the luminosity function at
redshifts higher than the peak in quasar evolution at $z\sim2$. In the
following section we will discuss both our model (\S\ref{lumfunc}), and the
HL98 model in light of the recently acquired data on quasars at high
redshifts.

\section{Comparison of Observed and Model Luminosity Functions}
\label{cp}

In this section we compute model luminosity functions at different
redshifts and compare them with the observed quasar luminosity
function. Figure~\ref{fig2} shows the observed luminosity function at
$z=0.1$, $1.0$, $2.3$, $3.0$, $4.0$, $z\sim4.3$ and $z\sim6.0$. Below
$z\sim4.3$, the data shown is from the summary by Pei~(1995; based on the
compilations of Hartwick \& Schade~1990). We also show the best fit
empirical luminosity function at $z\sim4.3$ from Fan et al.~(2001a) with a
vertical bar to denote the quoted error in the normalization. At $z\sim6$
the space density was inferred from Fan et al.~(2001b), and the error bar
includes uncertainty in the spectral index.  Comparisons of this data with
the halo merger rate model (\S\ref{comp}) and the HL98 model
(\S\ref{compHL}) are presented below.

\subsection{The Merger Rate Model}
\label{comp}

\begin{figure*}[htbp]
\epsscale{1.}
\plotone{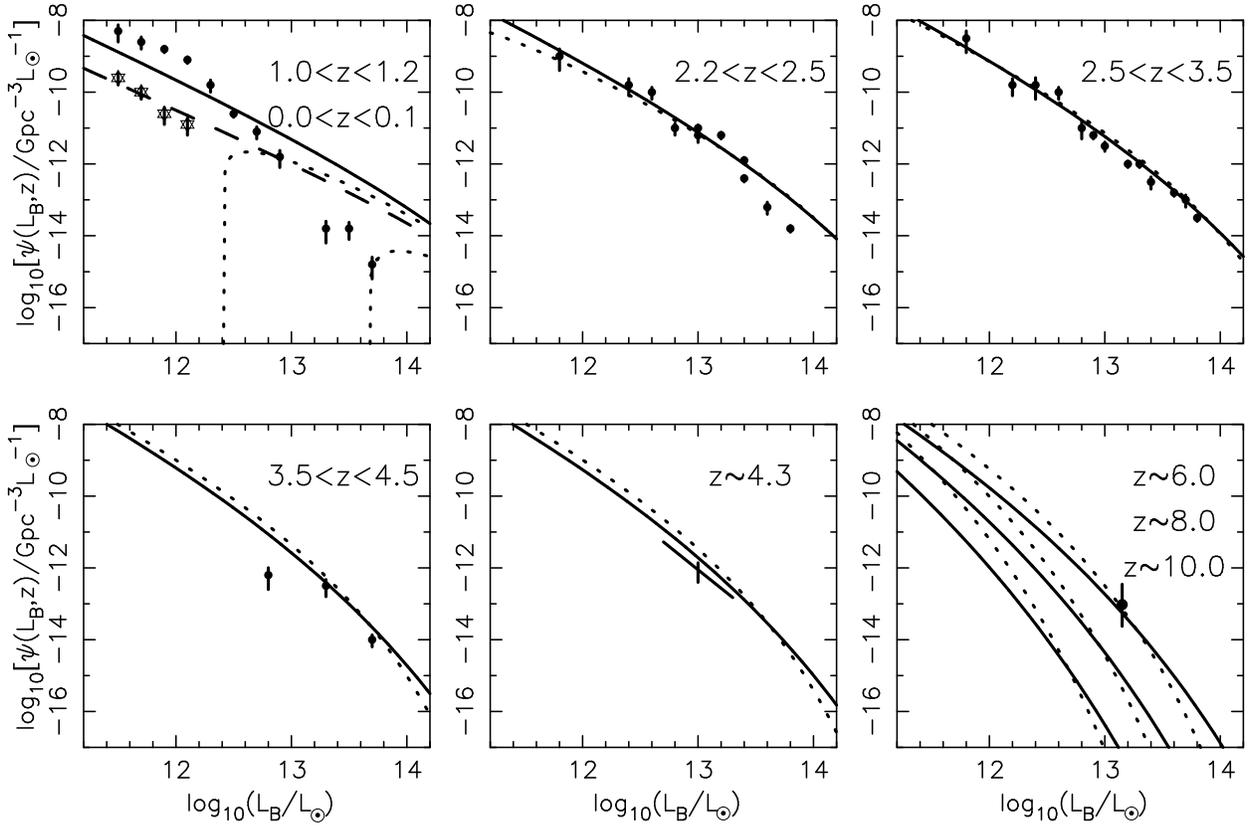}
\caption{\label{fig2} Comparison of the observed and model rest frame
B-band luminosity functions. 
The data at $z\la4$ is summarized in Pei~(1995). At $z\sim4.3$ and
$z\sim6.0$ the data is from Fan et al.~(2001a;2001b). In the $z\sim4.3$
panel, the diagonal line shows the best fit slope of $-2.58$ measured by
Fan et al.~(2001a), and the vertical bar shows the quoted uncertainty in
the normalization. In the lowest redshift panel, the dashed line and the
stars represent the merger model luminosity function and the observed
luminosity function data at $z\sim0.1$ respectively. The dotted lines show
the HL98 model ($\epsilon=10^{-3.1}$, $t_{\rm dc,0}=10^{5.0}$yr) and the 
solid lines are the merger model described in this paper ($\epsilon_{\rm 0}
=10^{-5.2}$, $\gamma=5$, $t_{\rm dc,0}=10^{6.3}$yr).}
\end{figure*}

Luminosity functions calculated from the halo merger rate model described
in \S\ref{lumfunc} are plotted as the solid lines in Figure~\ref{fig2} at
redshifts $z=0.1$, $1.0$, $2.3$, $3.0$, $4.0$, $z\sim4.3$ and
$z\sim6.0$. In addition, the lower right panel shows predicted luminosity
functions at $z=8.0$ and $z=10.0$. Values of $\epsilon_{\rm
o}=10^{-5.2}$, $\gamma=5$ and $t_{\rm dc,0}=10^{6.3}$ yr
yield a good fit to the data at all $z\ga2$. The parameters in the black
hole -- halo mass relation bear a striking resemblance to those in the relation
found by Ferrarese~(2002) for the local universe, which in our notation
correspond to $\epsilon_{\rm o}=10^{-5.1}$ and $\gamma=4.71$.  

The value of $t_{\rm dc,0}$ is related to the duty cycle for different
quasar luminosities and redshifts through equation~(\ref{dc}). The duty
cycle is plotted in Figure~\ref{fig5}. For a given halo mass, $t_{\rm dc}$
is longer at higher redshift. Steidel et al. (2002) estimated the lifetime
of bright quasar activity in a large statistical sample of Lyman-break
galaxies at $z\sim 3$ to be $\sim 10^7$ yr, which is surprisingly close
to our expected value of $t_{\rm dc}=10^{6.9}$ yr.
Above $z\sim3$ we find $t_{\rm dc}\sim10^{7}$ yr, comparable to the value 
inferred by comparing the local black hole 
density with the quasar luminosity function (e.g. Yu \&
Tremaine~2002). Note that while $t_{\rm dc}$ becomes a substantial fraction of
the age
of the universe at the highest redshifts considered, individual episodes of
quasar activity are shorter than $t_{\rm dc,0}$, which is an order of
magnitude smaller.

\begin{figure*}[htbp]
\epsscale{.7}
\plotone{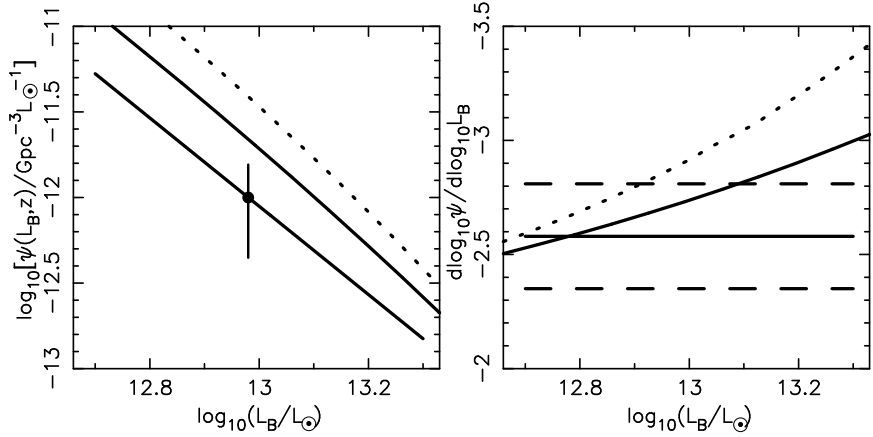}
\caption{\label{fig3} Detailed comparison of the observed rest frame
B-band quasar luminosity
function with the models at $z\sim4.3$. The left panel shows the luminosity
function in the region measured by Fan et al.~(2001a). The straight line
with the error bar shows the empirical luminosity function (best fit slope
of $\beta=-2.58$) and the quoted uncertainty in the normalization. The
upper solid and dotted lines show the luminosity functions for the halo
merger model ($\epsilon_{\rm 0}=10^{-5.2}$, $\gamma=5$, 
$t_{\rm dc,0}=10^{6.3}$yr) and for the HL98 model ($\epsilon=10^{-3.1}$, 
$t_{\rm dc,0}=10^{5.0}$yr), respectively. On the right panel we
show the derivatives of the model luminosity functions. The solid and
dotted lines show our model and the HL98 model , respectively. The empirical
best fit slope (solid) and the slopes $\pm1\sigma_\beta=\pm0.23$ away (dashed)
are shown for comparison.}
\end{figure*}

Two results from the comparison between model and data stand out. First,
the halo merger model reproduces the shallow slope for the $z\sim4.3$
quasars observed by Fan et al.~(2001a), and the space density of quasars at
$z\sim6.0$ measured by Fan et al.~(2001b). A detailed comparison between
the data and the model at $z\sim4.3$ is shown in Figure~\ref{fig3}. The
left hand panel shows the luminosity function in the region measured by Fan
et al.~(2001a). The straight line with the error bar represents the empirical
luminosity function (best fit slope of $\beta=-2.58$) and the uncertainty in 
its normalization. The upper solid line is the
luminosity function of the merger rate model. The theoretical luminosity
function slope is consistent with that of the observed quasars over the
measured luminosity range. This is shown more clearly on the right hand panel
of Figure~\ref{fig3} where the derivative of the model luminosity function
is shown in comparison with the empirical best fit slope (solid line) and
the slope $\pm1\sigma_\beta$ away (dashed lines). The model is consistent
within the quoted uncertainty over the measured luminosity range, particularly
at the fainter end. Our model
luminosity function is consistent with the observed lower limit on the
slope at $z=6.0$, which Fan et al.~(2000b) quote as $\beta_{\rm h}<3.9$
(95\%). However the gravitational lensing rate may be very high in this
sample (Wyithe \& Loeb~2002a,b). While the average slope of the luminosity
function cannot be changed significantly by lensing, we might easily be
misled by the magnification of 1 or 2 quasars in the limited current sample
of 4.  Furthermore, lensing might increase the apparent space density of
quasars by a factor of up to 2 (Wyithe \& Loeb~2002b).  The predicted
luminosity functions show that the slope of the bright end becomes
significantly steeper at redshifts higher than $z\sim6$ as expected from
the shape of the Press-Schechter~(1974) mass function.

The fit to the luminosity function discussed above has $\gamma=5$.
An equally good fit can be obtained for $\gamma=4$ and $\epsilon_{\rm
o}=10^{-4.5}$.  However this value of $\epsilon_{\rm o}$ is nearly an 
order of magnitude larger than observed (Ferrarese~2002), and the implied duty
cycle is nearly an order of magnitude too small (Yu \& Tremaine~2002; Steidel et
al. 2002). Hence, the complete set of observational constraints favors a
value of $\gamma=5$.  Ferrarese \& Merritt~(2002) find a relation between
$M_{\rm bh}$ and bulge velocity dispersion $\sigma_{\rm b}$
of $M_{\rm bh}\propto\sigma_{\rm b}^{4.72\pm0.36}$ while Tremaine et
al.~(2002, and references therein) find $M_{\rm bh}\propto\sigma_{\rm
b}^{4.02\pm0.32}$. Tremaine et al.~(2002) attribute this disagreement
primarily to the differences in the velocity dispersion measurements used
by the two groups.  Unfortunately, the highly uncertain nonlinear scaling
of $v_{\rm c}\propto \sigma_{\rm b}^{0.88\pm 0.17}$ (Ferrarese~2002) allows
the results of both groups to be consistent with our favored value of
$\gamma=5$ (which in turn translates to a mean slope of $M_{\rm
bh}\propto\sigma_{\rm b}^{4.4}$, in between the slopes of the two groups).

The upper left hand panel of Figure~\ref{fig2} indicates that the model
luminosity function provides a poorer fit to the data at redshifts
$z\la2$. The density of relatively faint [$L_{\rm B}\la10^{13}L_{\rm \odot}$ at
$z\sim1.0$ and $L_{\rm B}\la10^{12}L_{\rm \odot}$ at $z\sim0.1$ (dashed lines)]
quasars is reasonably reproduced, particularly at the lowest redshift
considered. The model also predicts a decline in the number of bright
quasars at low redshifts. On the other hand, the slope of the model
luminosity function is too shallow at the higher luminosities to explain
the data. A simple observational explanation for this phenomenon might be
the large drop in the average cosmic density of cold galactic gas from
$z\sim2$--$3$ to the present day (Storrie-Lombardi, McMahon \&
Irwin~1996). Furthermore, the inability of the gas to cool following the
epoch of group and X-ray cluster formation would also contribute to the
rapid decline in the luminosity function at low redshifts (Cavaliere \&
Vittorini 2000). Thus, while the merger rate predicts a decrease in the
space density of quasars at low redshifts, the more complicated low
redshift environments require additional modeling (e.g. Kauffmann \&
Haehnelt~2000) to reproduce the observed luminosity function shape.

\subsection{The Halo Formation Model}
\label{compHL}

We find that the HL98 model (\S\ref{modelhl}) requires\footnote{HL98 quote
$t_{\rm dc,0}=10^{5.8}$ yr, larger than our value of $t_{\rm
0}=10^5$. Haehnelt, Natarajan \& Rees~(1998) also found a
lower value $t_{\rm dc,0}$ in their model with constant $\epsilon$.}
best-fit values of $\epsilon=10^{-3.1}$ and $t_{\rm dc,0}=10^{5.0}$ yr. The
resulting luminosity function is shown as the dotted lines in
Figures~\ref{fig2} and \ref{fig3}. The HL98 model and the halo merger model
produce similar luminosity functions at $z\ga2$. However, the predicted duty 
cycle is too low by two orders of magnitude, and the value
of $\epsilon$ produced by this fit is too high by 2 orders of magnitude to be
consistent with observations.

\begin{figure*}[htbp]
\epsscale{.7}
\plotone{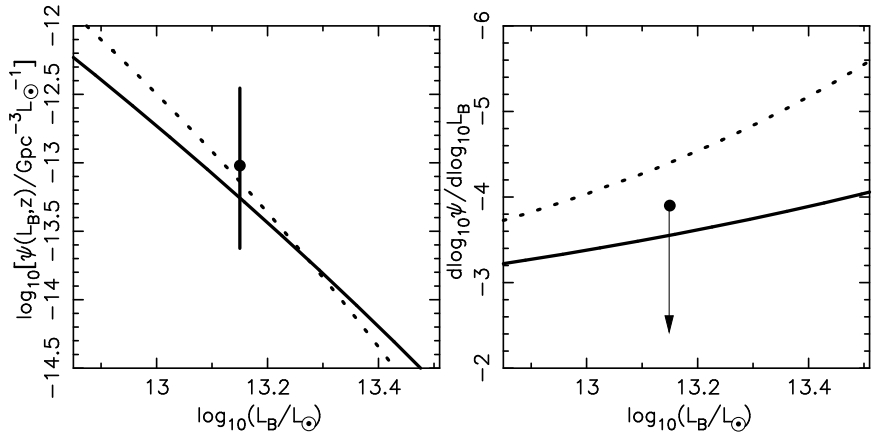}
\caption{\label{fig4} Detailed comparison of the observed  rest frame
B-band quasar luminosity
function with the models at $z\sim6.0$. The left panel shows the space
density of quasars measured by Fan et al.~(2001b). The solid and dotted
lines show the luminosity functions for the halo merger model 
($\epsilon_{\rm 0}=10^{-5.2}$, $\gamma=5$, $t_{\rm dc,0}=10^{6.3}$yr) 
and for the HL98 model ($\epsilon=10^{-3.1}$, $t_{\rm dc,0}=10^{5.0}$yr), 
respectively. On the right panel we show the derivatives of the
model luminosity functions.  The empirical lower limit (95\%) is shown for
comparison.}
\end{figure*}

\begin{figure*}[htbp]
\epsscale{.4}
\plotone{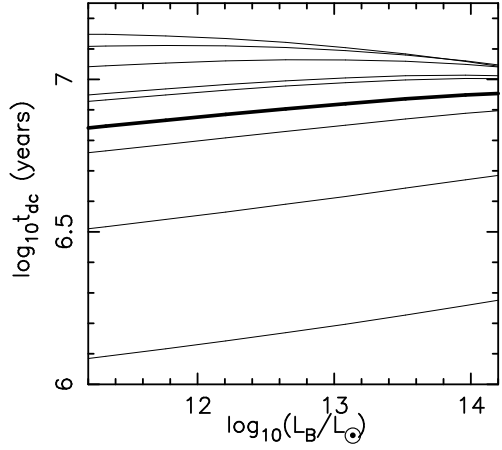}
\caption{\label{fig5} The duty cycle time $t_{\rm dc}$ of quasars with
luminosity $L_{\rm B}$ calculated from equation~(\ref{dc}) at redshifts $z=0.1$,
$1.0$, $2.3$, $3.0$, $4.0$, $4.3$, $6.0$, $8.0$ and $10.0$ (bottom to top, the 
thick line denotes the value at $z=3.0$ which should be compared with $\sim10^7$yr
found by Steidel et al.~(2002)). The duty cycle was calculated using our best fit 
value of $t_{\rm dc,0}=10^{6.3}$yr per merger.}
\end{figure*}

 Figures~\ref{fig3} and \ref{fig4} indicate that the HL98 model produces a 
somewhat less consistent normalization at $z\sim4.3$ and a steeper slope at 
$z\sim6.0$ when compared with the halo merger model. At low
redshifts the merger rate is higher than the collapse rate for halos in the
mass range of interest. In fact, the derivative of the Press-Schechter~(1974)
 mass function becomes negative and the model predicts a luminosity function with
a sharp lower cutoff in luminosity (in the implementation of HL98 the
smooth function, $f(t)$, results in a softer cutoff). This is seen in the
upper left panel of Figure~\ref{fig2}. If quasar activity was associated
solely with the halo formation rate, then only quasars residing within the
very largest halos would be observed at low redshift. 

\section{Rates of Multiple-Image Lensing}
\label{lens}
The fraction of quasars that are gravitationally lensed by foreground
galaxies is very sensitive to the underlying luminosity function due to the
potentially large magnification bias (Turner 1980; Turner, Ostriker \&
Gott~1984). At low redshifts ($z\la3$) the bright end of the luminosity
function has a universal slope of $-3.43$. However at $z\sim4.3$ Fan et
al.~(2001a) observed a slope of $-2.58$. Wyithe \& Loeb~(2002a,b) showed
that empirical extrapolations of the low redshift quasar luminosity
function to high redshifts yielded very different results for the
gravitational lensing rate if one were to adopt a bright end slope that is
$-3.43$ at all redshifts or a slope that is $-2.58$ above $z\sim3$.  Since
the halo merger model described in \S\ref{lumfunc} reproduces all measured
properties of the quasar luminosity function at high redshifts, we use it
to compute the magnification bias for multiple-image gravitational
lenses. The magnification bias can then be used to find the fraction of all
quasars (at different redshifts) that are multiply imaged. The lensing
fraction for the SDSS high redshift samples was discussed at length in
Wyithe \& Loeb~(2002a,b). We therefore give only a short summary here,
and present results for the lens fraction obtained with the new luminosity
function.

\begin{figure*}[htbp]
\epsscale{.7} \plotone{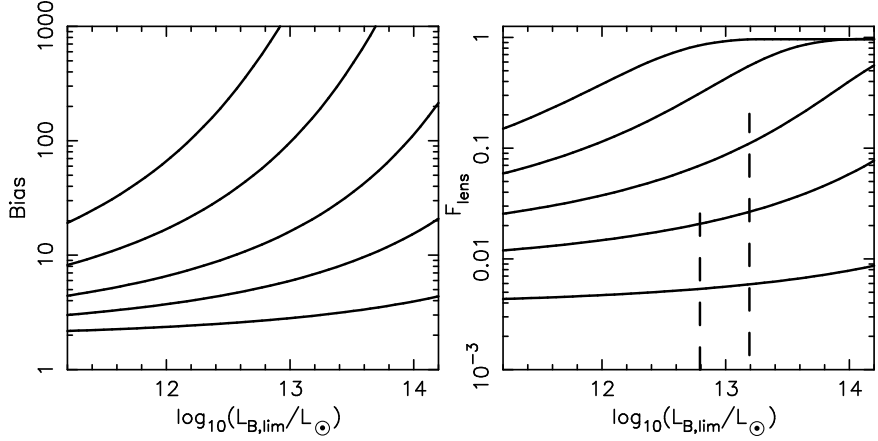}
\caption{\label{fig6} {\it Left:} The magnification bias as a function of
absolute luminosity limit for flux limited samples of quasars at $z=2.3$,
$4.3$, $6.0$, $8.0$ and $10.0$ (bottom to top). {\it Right:} The
multiple-image lensing rate for flux limited samples of quasars at $z=2.3$,
$4.3$, $6.0$, $8.0$ and $10.0$ (bottom to top). The left and right vertical dashed 
lines correspond to the absolute flux limit of the SDSS samples at $z\sim4.3$ and
$z\sim6.0$, respectively.}
\end{figure*}

The probability distribution of point source magnification by a singular,
spherical isothermal lens in terms of the sum of the magnifications of
multiple images, $\mu$, is $({dP}/{d\mu})={8}/{\mu^3}$ for $\mu>2$. The
bias factor for multiple imaging in a flux limited sample brighter than
(absolute luminosity) $L_{\rm B,lim}$ is therefore
\begin{equation}
B(z)=\frac{\int_2^\infty d\mu\frac{8}{\mu^3}\int_{\rm {L_{\rm
B,lim}}/{\mu}}^\infty dL'\Psi(L',z)}{\int_{\rm L_{\rm B,lim}}^\infty
dL'\Psi(L',z)}.
\end{equation}
This magnification bias is plotted as a function of the limiting (absolute)
luminosity in Figure~\ref{fig6}. Biases at redshifts of $z=2.3$, $z=4.3$
and $z=6.0$ are shown, as well as predictions for $z\sim8.0$ and
$z\sim10.0$ (bottom to top). The bias becomes very large at high redshifts
and for bright limiting magnitudes.

The resulting fraction of multiple images expected in the flux limited
sample is
\begin{equation}
F(L_{\rm B,lim},z) \approx \frac{B\tau_{\rm mult}}{B\tau_{\rm
mult}+(1-\tau_{\rm mult})}.
\end{equation} 
In this expression $\tau_{\rm mult}$ is the multiple image optical depth,
i.e. the source plane probability that a source will be multiply imaged for
the observer.  From Wyithe \& Loeb~(2002b) we find $\tau_{\rm mult}=0.0040$
for the SDSS quasars at $z\sim4.3$ and $\tau_{\rm mult}=0.0059$ for the
SDSS quasars at $z\sim6$. In addition, we find $\tau_{\rm mult}=0.0020$,
$\tau_{\rm mult}=0.0076$ and $\tau_{\rm mult}=0.0091$ (considering all
galaxies as potential lenses) at $z=2.3$, $z=8.0$ and $z=10.0$,
respectively.  The lens fraction $F(L_{\rm B,lim},z)$ is plotted on the right
hand panel of Figure~\ref{fig6} for $z=2.3$, $4.3$, $6.0$, $8.0$ and $10.0$
(bottom to top).  While at low redshifts the lensing fraction is less than
one percent, at higher redshifts the fraction could be much higher. The
vertical dashed lines correspond to the limiting absolute luminosities of
the SDSS quasar samples at $z\sim4.3$ (left) and $z\sim 6$ (right).  These
limits suggest lens fractions of $\sim2\%$ in the SDSS sample at $z\sim4.3$
and $\sim10\%$ in the SDSS sample at $z\ga5.7$. Note that the fractions at
$z\sim4.3$ and $z\sim6.0$ are consistent with the value obtained for the
flat luminosity function assumed in Wyithe \& Loeb~(2002b).  At even higher
redshifts the lens fraction approaches unity. The luminosity function at
$z\sim10$ predicts that quasars having luminosities brighter than a limit
that corresponds to the same co-moving space density as the SDSS quasars at
$z\sim6.0$ have a lens fraction of $F\sim0.5$.

\section{Summary and Discussion}
\label{conclusion}

We predict the quasar luminosity function under the hypothesis that quasar
activity is associated with galaxy mergers.  Galactic halos are assumed to
each host a black hole having a mass determined from a power-law relation
with halo circular velocity, $M_{\rm bh}\propto v_{\rm c}^\gamma$. Upon 
merger we assume that the gas from the
accreted halo is driven to the center, and that the quasar then shines at
its Eddington rate for a short period of time, proportional to the 
(smaller than 1/2) mass ratio between
the small and final halos in the merger. Our model has 3 free parameters but
at least the following 5 constraints: the luminosity function slope and
normalization at $z\sim 2.5$, the evolution of the luminosity function
slope and normalization, and the curvature of the luminosity function.  Our
model is simplified by the fact that the effect of reionization on the
luminosity function is limited to luminosities much fainter than those
detectable in current quasar surveys.

This very simple model reproduces all known properties of the luminosity
function above $z\sim2$, including the recently measured luminosity
function slope at $z\sim4.3$ (Fan et al.~2001a) and the space density of
quasars at $z\sim6.0$ (Fan et al.~2001b). The model suggests a
black hole -- halo mass relation, $M_{\rm bh}\propto v_{\rm c}^5$ consistent 
with that found by Ferrarese~(2002) at $z=0$, and a duty cycle of $\sim10^7$ yr
consistent with previous determinations
(Yu \& Tremaine~2002; Steidel et al. 2002). Using the derived luminosity
function we predict the resulting lensing rates for the highest redshift
samples. The recently published flux limited samples of SDSS quasars at
$z\sim4.3$ and $z\ga5.7$ are predicted to have multiple image fractions of
$\sim2\%$ and $\sim10\%$, respectively.

The model luminosity function computed using the halo merger rates is
consistent with the density of faint quasars at low redshifts, and predicts
a decline in the density of bright quasars towards low redshifts, but
predicts a luminosity function slope that is shallower than observed. As
has been noted previously, a more careful analysis (e.g. Kauffmann \&
Haehnelt~2000) is required to understand the evolution of the quasar
luminosity function at low redshifts. Nevertheless, the consistency of the
faint quasar normalization stands in contrast to models that associate
quasar activity with the halo formation rate, as those predict no faint
quasars at low redshifts.  Thus one is led to conclude that mergers are the
likely trigger for quasar activity at all redshifts.

The $M_{\rm bh}$--$v_{\rm c}$ relation that provides the best fit to the
data may also have a simple physical origin. First, consider dimensional
analysis. If we make the minimal assumption that the limiting luminosity of
a quasar is only a function of the halo circular velocity $v_{\rm c}$, then
the only dimensional parameters in the problem are $v_{\rm c}$ and $G$
(Newton's constant). The only combination of these parameters that has
dimensions of luminosity is $v_{\rm c}^5/G$ and so we conclude
\begin{equation} 
L_{\rm Edd} \propto {v_{\rm c}^5\over G}.
\end{equation} 
{\it What is the physics behind this relation?} The value of $v_{\rm
c}^5/G$ amounts to depositing the entire binding energy of a
self-gravitating system during its dynamical time (so that it does not have
time to adjust). The binding energy of a self-gravitating mass $M$ is $\sim
M v_{\rm c}^2$ and the dynamical time is $\sim r/v_{\rm c}$. The ratio
between these quantities is $\sim M v_{\rm c}^3/r$. Using the virial
relation $GM/r\sim v_{\rm c}^2$, we get the energy deposition rate that
would unbind a self-gravitating system on its dynamical time, $\sim v_{\rm
c}^5/G$. A quasar may therefore unbind the gas in the galaxy around
it\footnote{We implicitly assume that the central region of the galaxy
(such as the proto-bulge) is characterized by a circular velocity similar
to that of the galactic halo, even though it is dominated by baryons. This
assumption follows naturally from the nearly flat rotation curves of nearby
galaxies.}  if its power output is too large\footnote{Note that the power
output from the quasar may include both radiation and mechanical
energy. Substantial outflows are inferred to exist in broad absorption line
quasars and radio galaxies [see recent summary in Furlanetto \& Loeb
(2002), and references therein].}  (Silk \& Rees 1998; Ciotti \& 
Ostriker~2001). As the mass of the
black hole increases and the quasar's Eddington luminosity approaches this
limit, the feedback will generate a powerful galactic wind and terminate
the accretion that feeds the quasar.  Hence, the quasar phenomenon may be
self terminating (similarly to the formation of a proto-star).  Of course,
if some of the quasar energy escapes (due to incomplete absorption, partial
covering factor or efficient cooling by the surrounding gas), then the
feedback-limited quasar luminosity would be higher than $v_c^5/G$.  As long
as the feedback from all quasars encounters self-similar conditions in
different halos, we may write
\begin{equation} 
\label{led}
L_{\rm Edd}= \beta_\epsilon {v_{\rm c}^5\over G} ,
\label{eq:beta}
\end{equation} 
where $\beta_\epsilon$ is a constant that is related to the inferred value
of $\epsilon_{\rm o}$.  Equations (\ref{vc}), (\ref{eps}), (\ref{led}) and
the best-fit value of $\epsilon_{\rm o}\sim 10^{-5.2}$ together with
the relation $L_{\rm Edd}=1.4\times 10^{38}(M_{\rm bh}/M_\odot)~{\rm
erg~s^{-1}}$, yield $\beta_\epsilon\approx 50$.
We have shown that the assumption of a constant $\beta_\epsilon$ in 
equation~(\ref{eq:beta}) leads to a luminosity function that describes the 
data over a wide range of redshifts. The constancy of $\beta_\epsilon$ (which 
is also expected from dimensional analysis) implies
that feedback may indeed be the mechanism that regulates the growth of
supermassive black holes in galactic potential wells.  The similarity
between our deduced values of $\gamma=5$ and $\epsilon_{\rm o}= 10^{-5.2}$
at high redshifts and the values inferred for remnant black holes in the
local universe (Ferrarese 2002; Tremaine et al. 2002), provides a strong
testimony to this effect.

\acknowledgements This work was supported in part by NSF grants
AST-9900877, AST-0071019 for AL. JSBW is supported by a Hubble Fellowship
grant from the Space Telescope Science Institute, which is operated by the
Association of Universities for Research in Astronomy, Inc., under NASA
contract NAS 5-26555.

\newpage

\end{document}